\documentclass[11pt]{elsarticle}

\pdfoutput=1

\makeatletter
\def\ps@pprintTitle{%
	\let\@oddhead\@empty
	\let\@evenhead\@empty
	\let\@oddfoot\@empty
	\let\@evenfoot\@oddfoot
}
\makeatother
\usepackage{float}
\usepackage{url}
\usepackage{breakurl}
\usepackage{amsmath}
\usepackage[breaklinks,
colorlinks = true,
linkcolor = blue,
urlcolor  = blue,
citecolor = blue,
anchorcolor = blue]{hyperref}

\usepackage{graphicx}
\usepackage[margin=1.25in]{geometry}
\usepackage[usenames,dvipsnames]{color}

\newcommand\acp{\begin{center}
		\rule[-0.2in]{\hsize}{0.01in}\\\rule{\hsize}{0.01in}\\
		\vskip 0.1in Submitted to the  Proceedings\\ 
		of the African Conference on Fundamental and Applied Physics
		\vskip 0.05in
		{\it Second Edition, ACP2021, March 7--11, 2022 --- Virtual Event}\\
		\rule{\hsize}{0.01in}\\\rule[+0.2in]{\hsize}{0.01in} \\
\end{center}}

\usepackage[firstpage=true]{background}
\backgroundsetup{contents={\parbox{6.5in}{\acp}}, scale=1,placement=top,opacity=1,color=black,position={3.25in,1.2in}}

\usepackage{fancyhdr}
\fancypagestyle{plain}{%
	\fancyhf{}%
	\fancyhead[C]{}
	\fancyfoot[C]{\thepage}
}

\fancypagestyle{empty}{%
	\fancyhf{}%
	\fancyhead[C]{{\it ACP2021, March 7--11, 2022 --- Virtual Edition}}
	\fancyfoot[C]{\thepage}
}
\begin{document}
	
	\begin{frontmatter}
		
		
		\title{Search of new resonances decaying into top quark pairs in the lepton+jet final state in proton-proton collisions at $\sqrt{s}$ = 13 TeV with the ATLAS detector}
		
		\author[add1]{Badr-eddine Ngair\corref{cor1}}
		\ead{Badr-eddine.ngair@cern.ch}
		\author[add1]{Farida Fassi}
		
		\cortext[cor1]{Corresponding Author}
		
		\address[add1]{Mohammed V University in Rabat, Faculty of Science, Morocco}
		
		\begin{abstract}
			\noindent 
	\small	A search for resonances produced in 13 TeV proton-proton collisions and decaying into top-quark pairs is presented. In this study events where the top-quark decay produces a single isolated charged lepton, missing transverse momentum and jet activity compatible with a hadronic top-quark decay recorded with the ATLAS detector at the Large Hadron Collider (LHC) are considered. We investigate the observed invariant mass spectrum in a model- independent approach to seek for any significant deviation from the Standard Model (SM) background expectation. The Matrix Method was used to estimate the QCD multi-jet background, which has large statistical and systematic uncertainties when modelled using Monte Carlo techniques. We have performed general searches for new resonances for a specific benchmark models: $Z^{'}_{TC2}$ boson, Kaluza–Kein gluon and Kaluza–Klein graviton that decay into top-quark pairs. Taking into account all uncertainties, results are in line with SM expectations. A synopsis of the results followed by an explanation of key findings will be presented.
	\end{abstract}
		

		
	\end{frontmatter}

	\section{Introduction}
	\label{sec:intro}
	\noindent
	LHC \cite{Evans:2008zzb} is the largest and the highest energy particle collider ever built to date in the world, located on the border of France and Switzerland near Geneva. The collider tunnel consist of four main experiments installed in a circular circumference of 27 km and 100 m underground. The ATLAS experiment \cite{ATLAS:2008xda} is a multi-purpose particle detector studying the signatures left
	by particles emerging from the proton-proton (pp) collisions.
	This proceeding presents a comprehensive overview of the sensitivity of the ATLAS detector to new resonances in the $t\bar{t}$ invariant mass spectrum. Several benchmark scenarios are explored in this search, such as the spin-1 $Z^{'}_{TC2}$ boson with 1.2\%
	 width predicted by the topcolour-assisted technicolour (TC2) model \cite{Hill_1995} simulated using the Pythia 8 generator default settings, Kaluza-Klein gluon (gKK) \cite{Lillie_2007} with width of 30\% generated using Pythia 8, and Spin-2 Kaluza-Klein graviton (GKK) \cite{Agashe_2007} signal generated using Madgraph 5 LO interfaced with Pythia 8. 
	 The study is done using 36.1 $fb^{-1}$ data collected at a centre-of-mass energy of 13 TeV using the ATLAS detector at the LHC.
		\section{Search for $t\bar{t}$ resonances in the lepton+jet final state}
\begin{figure}[H]

				\centering
				\includegraphics[width=6.5cm,height=4cm]{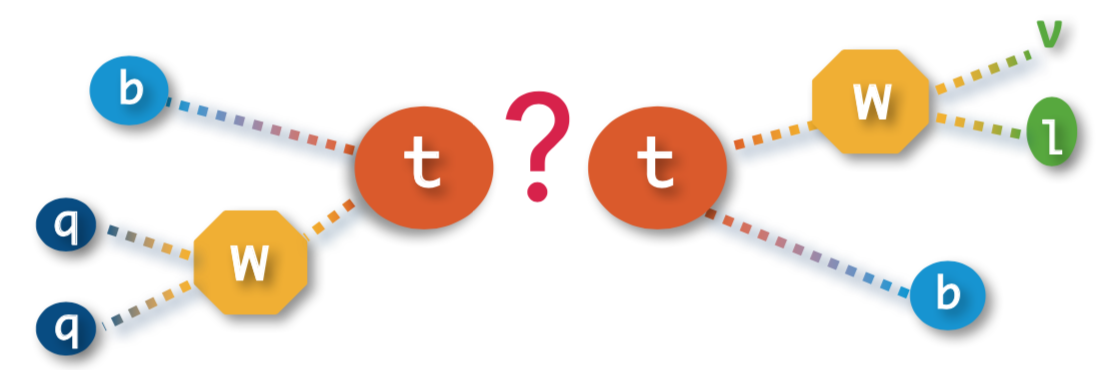}
		
			\caption{\small 	A preventative diagram of the searched signal process where one of the top quarks decay hadronically and the other one decays leptonically.}
			\label{fig:fey}
		\end{figure}

		The search for new physics that produces resonances in the top-antitop ($t\bar{t}$) invariant mass spectrum \cite{ATLAS:2018rvc} is carried out in the semi-leptonic ($\ell$+ jets) decay channel. 
		A preventative diagram of this process is shown in Figure \ref{fig:fey}. 
		
		The signal is a significant deviation (“bump”) from the $t\bar{t}$ mass spectrum expected by the SM.
		 The selection starts by requiring events with exactly one isolated charged lepton (electron or muon) with transverse momentum ($p_{T} $) greater than 25 GeV, large missing transverse momentum (whose magnitude is denoted by $E_{T}^{miss}$), and hadronic jets. Candidate events are required to contain at least one b-tagged jet and can be further split into four categories based on the number of the b-tagged jets as follows: Category 0 considers only events with no b-tagged jet associated to the hadronic or leptonic-top candidates, while categories 1 and 2 consider events with only the leptonic and hadronic top candidate that have a matched b-tagged jet, respectively. Category 3 contains events with both the hadronic-top and the leptonic-top candidates that have a matched b-tagged jet. 
		 The analysis is optimized to adapt to different scenarios: boosted and resolved. The three types of jets considered in this analysis are reconstructed using the $Anti-k_{t}$ algorithm \cite{Cacc}: the small-R jet are built from topological clusters formed by energy deposited in the hadronic calorimeter, satisfying $p_{T} >$ 25 GeV, $|\eta| <$ 2.5 and a radius parameter R=0.4. The large-R jets with radius parameter R = 1.0 are required to have $p_{T} >$ 300 GeV and $|\eta| <$ 2.0. The track-jets are reconstructed from charged tracks, with R=0.2, $p_{T} >$ 10 GeV and
	$|\eta| <$ 2.5. Events in the boosted channel should contain at least one small-R jet ($J_{sel}$) with $\Delta R(J_{sel},lepton) < $ 1.5 ($\Delta R = \sqrt{  \Delta\eta^{2} +  \Delta\phi^{2}}$). The
	large-R jet is required to satisfy an azimuthal angle cut from both the lepton ($\Delta \phi>$ 2.3 rad) and from $J_{sel}$ ($\Delta \phi>$  1.5 rad). Events that fail the selection criteria of the boosted channel are tested against the selection criteria of the resolved topology where the events must contain at least four Small-R jets with $p_{T} >$ 25 GeV and only events with $log_{10}(\chi2) <$ 0.9 are kept. To suppress the QCD multijets events, cuts on the $E_{T}^{miss} >$ 20 GeV and $E_{T}^{miss} + m_{T}^{W}$\footnote{$m_{T}^{w}= \sqrt{2.p_{T,\ell}.E_{T}^{miss}(1-cos(\Delta \Phi)}$ is the W boson transverse mass, where $p_{T,\ell}$ is the transverse momentum of the lepton and $\Delta \Phi $ is the angular distance between lepton and the $E_{T}^{miss}$.} 
	$>$ 60 GeV are applied.
	\section{$t\bar{t}$  invariant mass reconstruction}
	After all the event selections, mass of the $t\bar{t}$ system ($m_{t\bar{t}}^{reco}$) is formed by combining the four-vector of the reconstructed objects. For the boosted-topology, the $m_{t\bar{t}}^{reco}$ is reconstructed from the four-momentum sum of the selected jet, the lepton, the neutrino and the leading large-R jet. For the resolved-topology, a $\chi2$ minimization algorithm is used to select and assign the small-R jets to the leptonically and hadronically decaying top quarks. The $\chi2$ function is constructed using the constraints from the expected top quark and W boson masses, also the top pair $p_{T}$ balance as shown in Equation \ref{eq:eq1}.

\begin{equation}
\begin{split}
	\chi^2 = & \left[ \frac{m_{jj}-m_W}{\sigma_W} \right]^2 + 
	\left[ \frac{m_{jjb}-m_{jj}-m_{th-W}}{\sigma_{th-W}} \right]^2 + \left[ \frac{(p_{\mathrm T,jjb}-p_{\mathrm T,j\ell\nu}) - 
		(p_{\mathrm T,th}-p_{\mathrm T,t\ell})}{\sigma_{diffp\mathrm T}}  \right]^2 
 \\
& 	+	\left[ \frac{m_{j\ell\nu}-m_{tl}}{\sigma_{t\ell}} \right]^2 
\end{split}
	\label{eq:eq1}
\end{equation}

\begin{figure}[H]
		\centering
		\begin{minipage}{.3\textwidth}
			\centering
			\includegraphics[width=5.2cm,height=5.3cm]{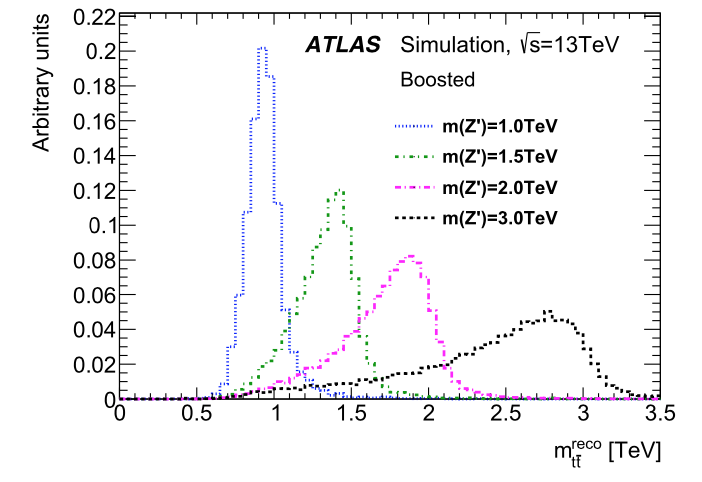}
		\end{minipage}\hfill
		\begin{minipage}{.33\textwidth}
			\centering
			\includegraphics[width=5.5cm,height=5.3cm]{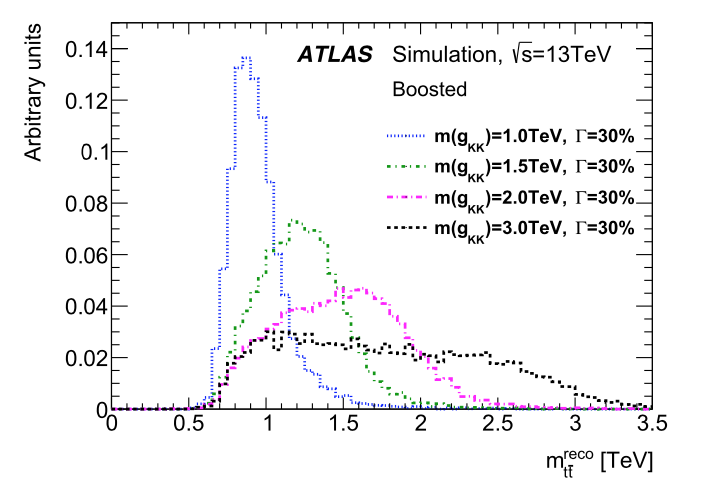}
		\end{minipage}\hfill
		\begin{minipage}{.3\textwidth}
			\centering
			\includegraphics[width=5.5cm,height=5.3cm]{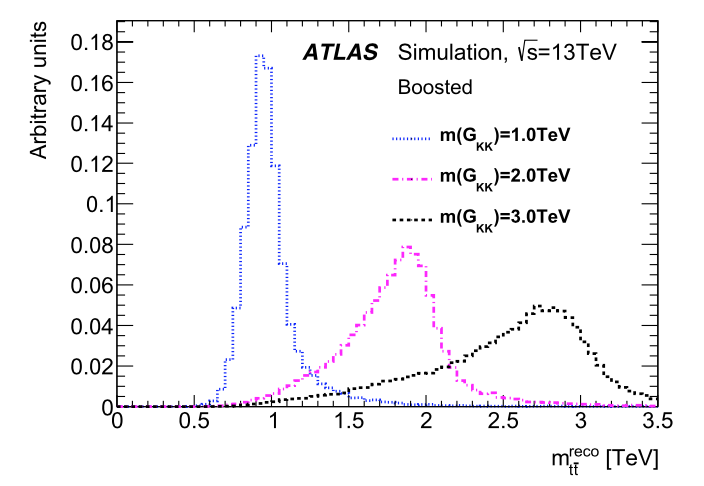}
		\end{minipage}
		\caption{\small $m_{t\bar{t}}$ distributions for the three signal models: $Z^{'}_{TC2}$ (left), $g_{KK}$ (middle) and $G_{KK}$ (right) for the events satisfying the resolved (top) and boosted (bottom) selections \cite{ATLAS:2018rvc}.}
		\label{fig:mttreco}
	\end{figure}

	The $t\bar{t}$ invariant mass distributions for several signal masses  are shown in Figure \ref{fig:mttreco}, where all events that pass the resolved and the boosted selection criteria are considered.
		\section{{Results}}
\begin{figure}[H]
		\centering
		\begin{minipage}{0.45\textwidth}
			\centering
			\includegraphics[width=6cm,height=5.3cm]{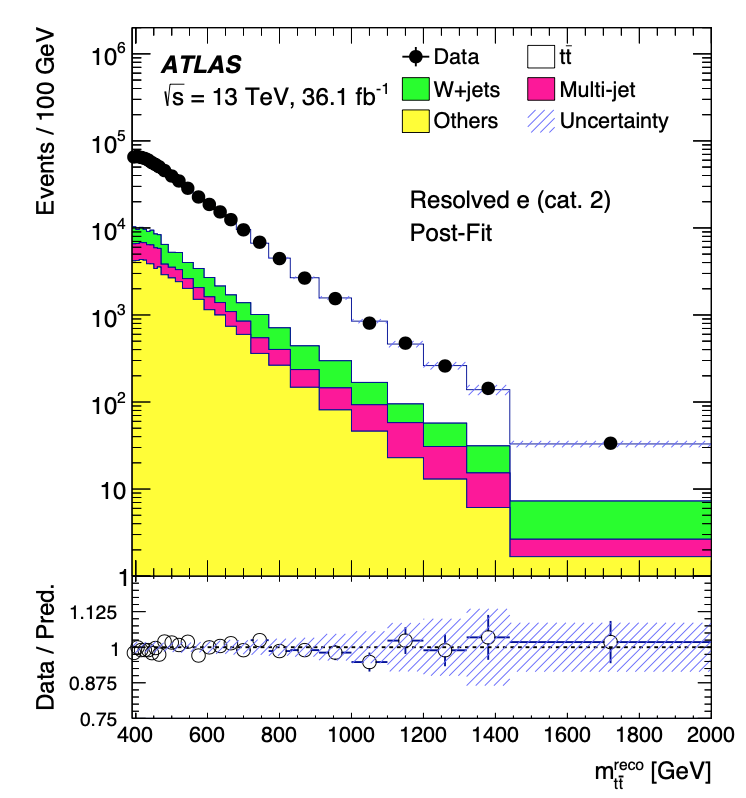}
		\end{minipage}\hfill
		\begin{minipage}{0.45\textwidth}
			\centering
			\includegraphics[width=6cm,height=5.3cm]{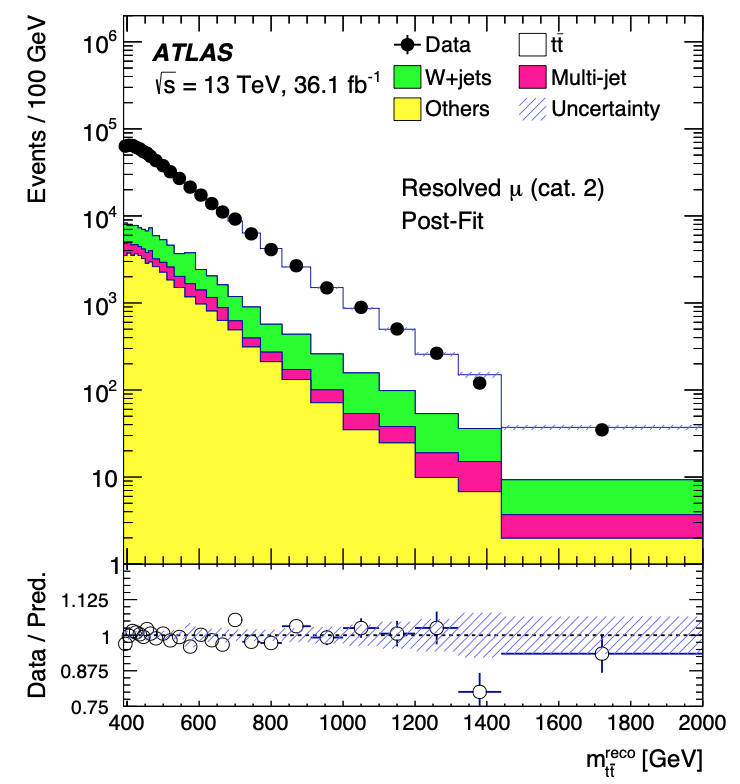} 
		\end{minipage}
		\caption{\small $m_{t\bar{t}}$ distributions after the profiling of the nuisance parameters in the boosted regime (top) and resolved regime (bottom) for the electron (left) and muon (right) channels \cite{ATLAS:2018rvc}.}
		\label{fig:mtt}
	\end{figure}

	The discriminating variables employed to look for massive resonances are the  $m_{t\bar{t}}^{reco}$ spectra from the two decay topologies (Boosted and Resolved). The observed data agrees well with the SM expectation within the covered systematic uncertainties associated to physics objects identification, reconstruction and calibration, jets energy scales and resolutions, as well as the theoretical uncertainties affecting $t\bar{t}$ irreducible background. A statistical fit is performed to the invariant mass of the $t\bar{t}$ system using an hypothesis testing tool, a so called BumpHunter (as shown in Figure \ref{fig:mtt}) with all nuisance parameters are considered in the profiling. The $t\bar{t}$ modelling and the large-R jet energy scale uncertainties are the major theoretical and experimental systematic uncertainties respectively. Based on the fit, using several hypotheses for the mass and width, it is possible to set observed  and expected limits on the cross-section times branching ratio for the different tested BSM models.
		\begin{figure}[H]
				\hspace{-0.5cm} 	
					\vspace{0.4cm} 
		\centering
		\begin{minipage}{.3\textwidth}
			\centering
			\includegraphics[width=5.3cm,height=5.3cm]{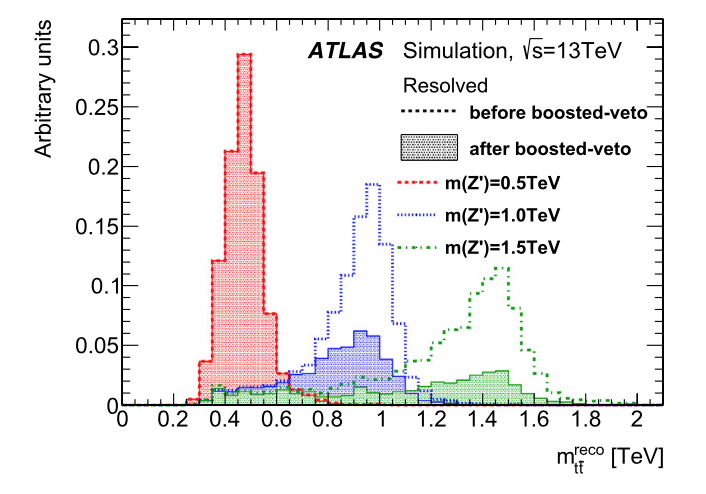}
			\label{fig:test1}
		\end{minipage}\hfill
		\begin{minipage}{.33\textwidth}
			\centering
			\includegraphics[width=5.5cm,height=5.3cm]{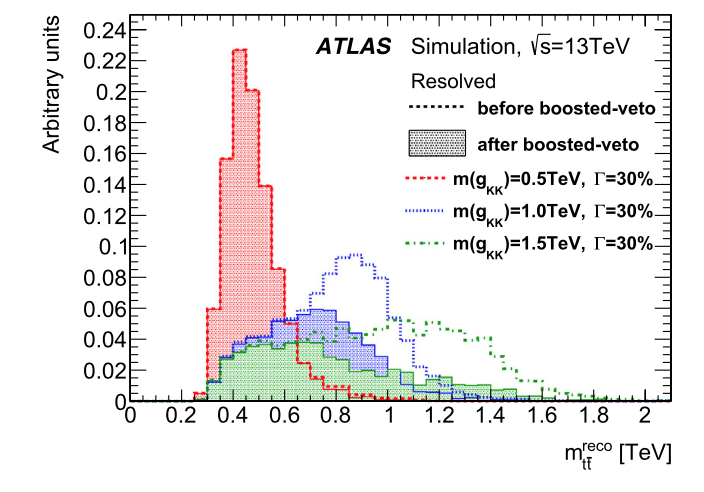}
			\label{fig:test2}
		\end{minipage}\hfill
		\begin{minipage}{.3\textwidth}
			\centering
			\includegraphics[width=5.5cm,height=5.3cm]{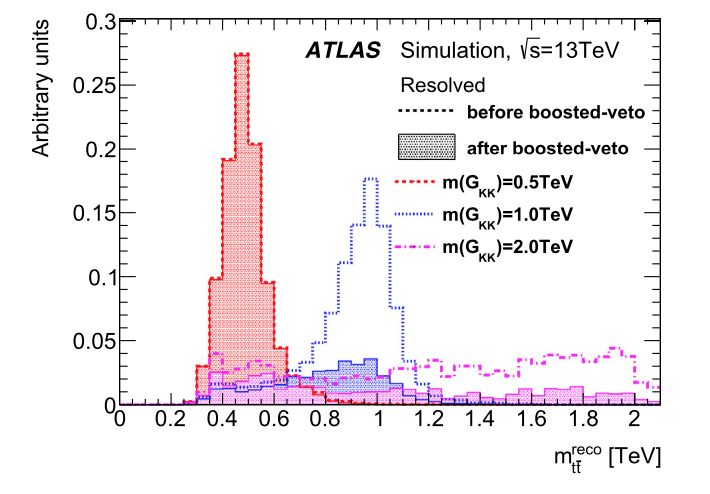}
			\label{fig:test3}
		\end{minipage}
		\end{figure}
		\hspace{-0.5cm} 	
		\vspace{-1.45cm} 

		\begin{figure}[H]
	
		\hspace{-0.5cm} 	
			\vspace{-0.3cm} 		
		\centering
		\begin{minipage}{0.45\textwidth}
			\centering
			\includegraphics[width=6cm,height=5.3cm]{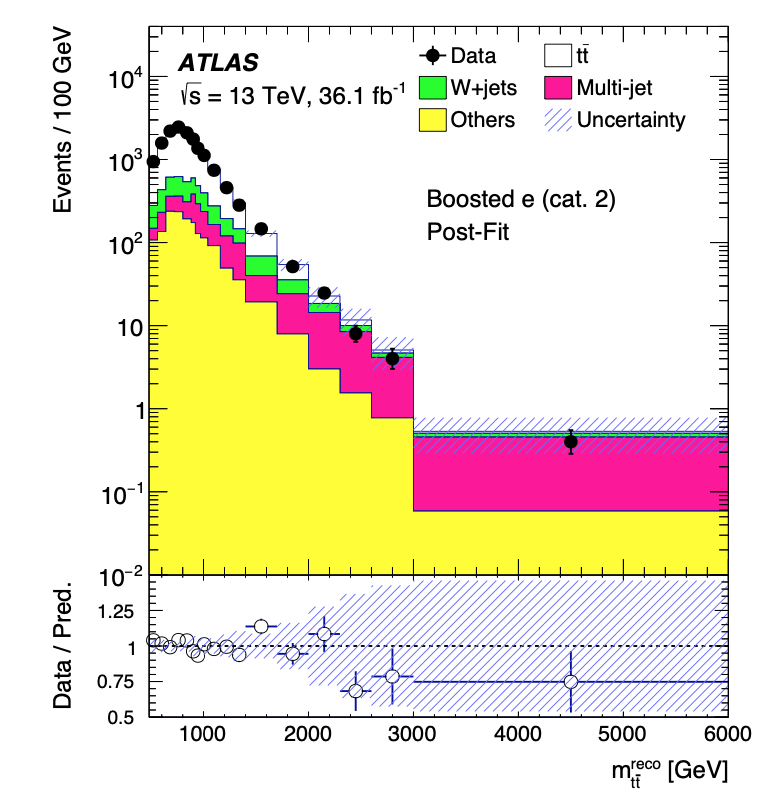} 
		\end{minipage}\hfill
		\begin{minipage}{0.45\textwidth}
			\centering
			\includegraphics[width=6cm,height=5.3cm]{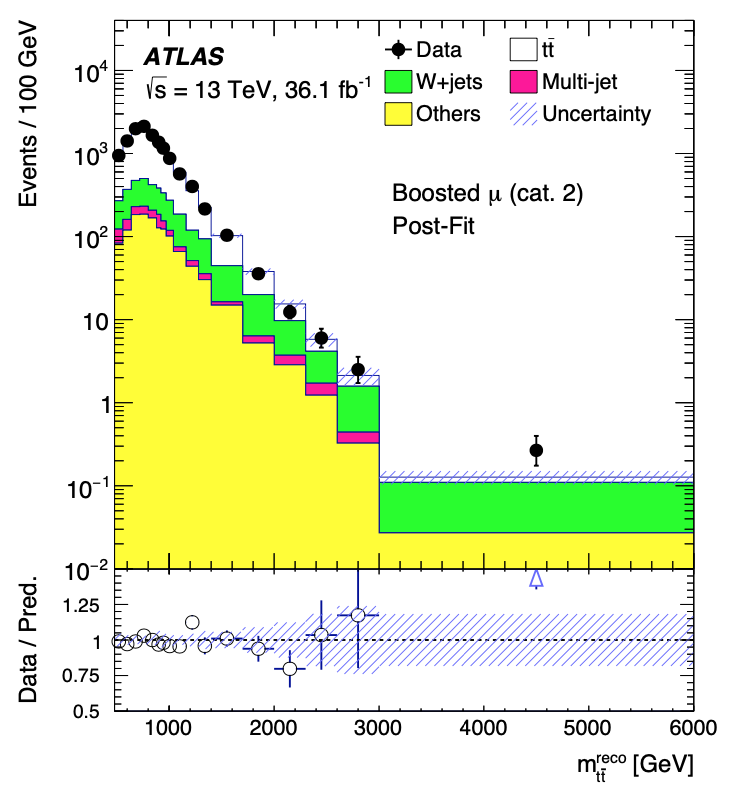} 
		\end{minipage}
	\end{figure}
	\hspace{-0.5cm} 	
	\vspace{-0.8cm} 
	
\begin{figure}[H]
		\centering
		\begin{minipage}{.3\textwidth}
			\centering
			\includegraphics[width=5.3cm,height=5.3cm]{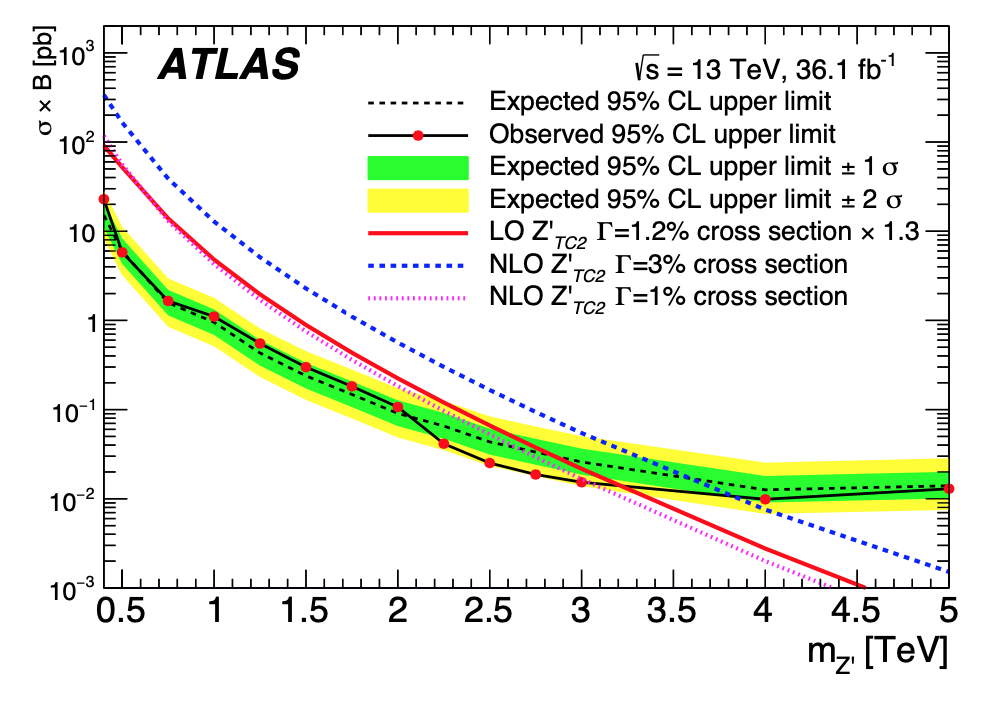}
		\end{minipage}\hfill
		\begin{minipage}{.33\textwidth}
			\centering
			\includegraphics[width=5.5cm,height=5.75cm]{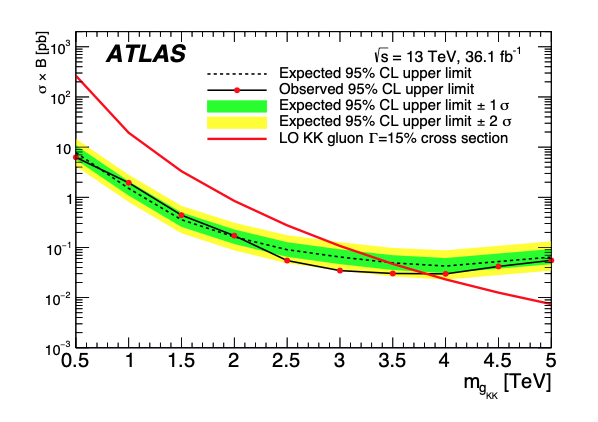}
		\end{minipage}\hfill
		\begin{minipage}{.3\textwidth}
			\centering
			\includegraphics[width=5.3cm,height=5.3cm]{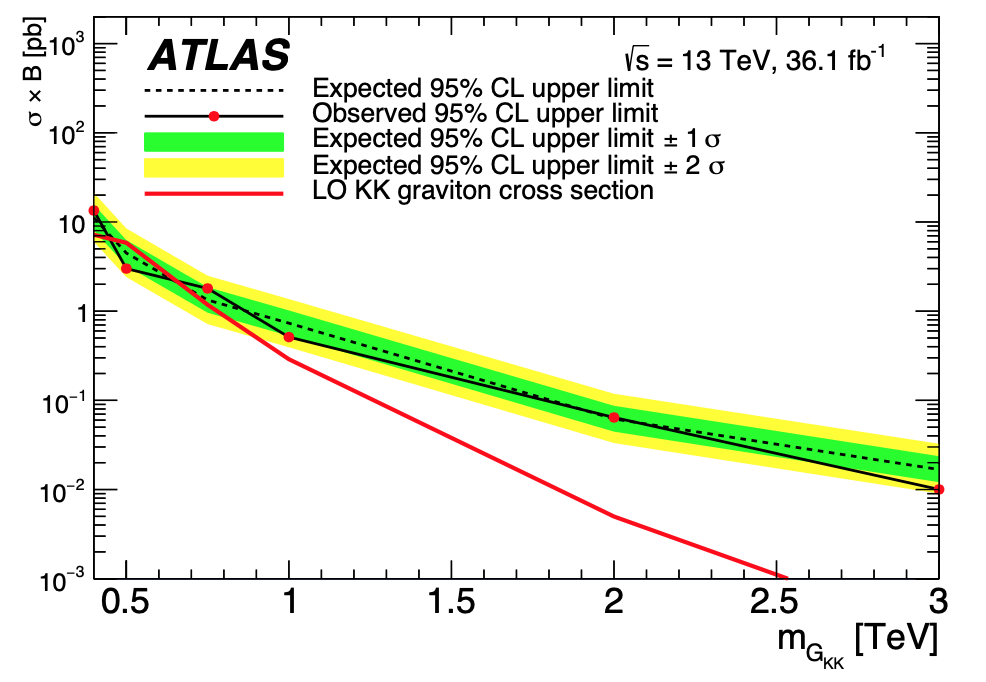}
		\end{minipage}
		\caption{\small Cross-section x branching ratio 95\% CL upper limits on the $Z^{'}_{TC2}$ signal (left), $g_{kk}$ (middle), and $G_{KK}$ (right) \cite{ATLAS:2018rvc}.}
			\label{fig:limit}
	\end{figure}

A sensitivity analysis to BSM resonances over the mass range from 0.4 TeV to 5.0 TeV is performed. The resulting cross-section times branching ratio 95 \% CL upper limits on the mass; a Topcolour-assisted
	Technicolour, $Z'_{TC2}$ , with a width of 1\% is excluded for masses  $Z'_{TC2}$ $<$ 3.0 TeV. The search allowed to exclude at 95\% C.L. Kaluza–Klein (KK) gravitons in the range 0.45 TeV $<$ $m_{G_{KK}}$ $<$ 0.65 TeV , while a KK gluon of width 15\% is excluded for  $m_{g_{KK}}$ $< $  3.8 TeV as shown in Figure \ref{fig:limit}.

	\vspace{-0.5cm} 
		\hspace{-0.5cm} 	
	\section{{conclusion}}
	 	In this proceeding, a search for new, heavy resonances decaying to a top-antitop pair in the $\ell+jets$ final state has been presented. The search was performed using 36.1 $fb^{-1}$ of data from pp collisions collected by the ATLAS detector at a centre-of-mass energy of $\sqrt{s}$ = 13 TeV. No evidence for the existence of new top-antitop quark resonances has been observed, and upper limits on the production cross section are set for the studied benchmark models. The new results using full Run-2 dataset might be available before summer 2022, where an improvement of sensitivity is expected. 

\small
\bibliographystyle{elsarticle-num}
\bibliography{myreferences} 

\begin{thebibliography}{1}
\expandafter\ifx\csname url\endcsname\relax
  \def\url#1{\texttt{#1}}\fi
\expandafter\ifx\csname urlprefix\endcsname\relax\def\urlprefix{URL }\fi
\expandafter\ifx\csname href\endcsname\relax
  \def\href#1#2{#2} \def\path#1{#1}\fi

\bibitem{Evans:2008zzb}
{LHC Machine}, JINST 3 (2008) S08001.
\newblock \href {https://doi.org/10.1088/1748-0221/3/08/S08001}
  {\path{doi:10.1088/1748-0221/3/08/S08001}}.

\bibitem{ATLAS:2008xda}
G.~Aad, et~al., {The ATLAS Experiment at the CERN Large Hadron Collider}, JINST
  3 (2008) S08003.
\newblock \href {https://doi.org/10.1088/1748-0221/3/08/S08003}
  {\path{doi:10.1088/1748-0221/3/08/S08003}}.

\bibitem{Hill_1995}
C.~T. Hill, \href{https://doi.org/10.1016%2F0370-2693%2894%2901660-5}{{Topcolor
  assisted technicolor}}, Physics Letters B 345~(4) (1995) 483--489.
\newline\urlprefix\url{https://doi.org/10.1016%2F0370-2693%2894%2901660-5}

\bibitem{Lillie_2007}
B.~Lillie, L.~Randall, L.-T. Wang, {The Bulk {RS} {KK}-gluon at the {LHC}},
  Journal of High Energy Physics 2007~(09) (2007) 074--074.
\newblock \href {https://doi.org/10.1088/1126-6708/2007/09/074}
  {\path{doi:10.1088/1126-6708/2007/09/074}}.

\bibitem{Agashe_2007}
K.~Agashe, H.~Davoudiasl, G.~Perez, A.~Soni, {Warped gravitons at the {CERN}
  {LHC} and beyond}, Physical Review D 76~(3) (aug 2007).
\newblock \href {https://doi.org/10.1103/physrevd.76.036006}
  {\path{doi:10.1103/physrevd.76.036006}}.

\bibitem{ATLAS:2018rvc}
M.~Aaboud, et~al., {Search for heavy particles decaying into top-quark pairs
  using lepton-plus-jets events in proton\textendash{}proton collisions at
  $\sqrt{s} = 13$ TeV with the ATLAS detector}, Eur. Phys. J. C 78~(7) (2018)
  565.
\newblock \href {http://arxiv.org/abs/1804.10823} {\path{arXiv:1804.10823}},
  \href {https://doi.org/10.1140/epjc/s10052-018-5995-6}
  {\path{doi:10.1140/epjc/s10052-018-5995-6}}.

\bibitem{Cacc}
M.~Cacciari, G.~P. Salam, G.~Soyez, {The anti-k-t jet clustering algorithm},
  Journal of High Energy Physics 2008 (2008) 063--063.
\newblock \href {https://doi.org/10.1088/1126-6708/2008/04/063}
  {\path{doi:10.1088/1126-6708/2008/04/063}}.

\end{thebibliography}


\begin{thebibliography}{10}
\expandafter\ifx\csname url\endcsname\relax
  \def\url#1{\texttt{#1}}\fi
\expandafter\ifx\csname urlprefix\endcsname\relax\def\urlprefix{URL }\fi
\expandafter\ifx\csname href\endcsname\relax
  \def\href#1#2{#2} \def\path#1{#1}\fi

\bibitem{ASP2021-reports}
{Kétévi A. Assamagan, Bobby Acharya, Temitope Adenuga, Mohamed Chabab,
  Kenneth Cecire, Simon H. Connell, Anne E. Dabrowski, Christine Darve, Farida
  Fassi, Jonathan R. Ellis, Fernando Ferroni, Mounia Laassiri, Steve G.
  Muanza}, {Activity report of the African School of Physics, 2019-2021},
  {arXiv:2109.00509} ({2019--2021}).
\newblock \href {https://doi.org/https://doi.org/10.48550/arXiv.2109.00509}
  {\path{doi:https://doi.org/10.48550/arXiv.2109.00509}}.

\bibitem{ASP}
{B. S. Acharya, K. A. Assamagan, A. E. Dabrowski, C. Darve, J. Ellis, S.
  Muanza}, {The African School of Physics},
  \url{https://www.africanschoolofphysics.org/}.

\bibitem{ASP-reports}
{Activity reports of African School of Physics},
  \url{http://africanschoolofphysics.web.cern.ch/2010/asp2010.pdf,
  https://africanschoolofphysics.web.cern.ch/asp2012/asp2012_final.pdf,
  https://www.africanschoolofphysics.org/wp-content/uploads/2014/11/asp2014.pdf,
  https://www.africanschoolofphysics.org/wp-content/uploads/2019/08/ASP2016-FinalReport.pdf,
  https://www.africanschoolofphysics.org/wp-content/uploads/2019/08/ASP2018.pdf}
  (2010-2018).

\bibitem{ASP-COVID}
{Kossi Amouzouvi, Kétévi A. Assamagan, Somiéalo Azote, Simon H. Connell,
  Jean Baptiste Fankam Fankam, Fenosoa Fanomezana, Aluwani Guga, Cyrille E.
  Haliya, Toivo S. Mabote, Francisco Fenias Macucule, Dephney Mathebula,
  Azwinndini Muronga, Kondwani C. C. Mwale, Ann Njeri, Ebode F. Onyie, Laza
  Rakotondravohitra, George Zimba}, {A model of COVID-19 pandemic evolution in
  African countries}, Scientific African vol. 14 ({2021}) e00987.
\newblock \href {http://arxiv.org/abs/2104.09675} {\path{arXiv:2104.09675}},
  \href {https://doi.org/https://doi.org/10.1016/j.sciaf.2021.e00987}
  {\path{doi:https://doi.org/10.1016/j.sciaf.2021.e00987}}.

\bibitem{asp2018}
{B. S. Acharya, K. A. Assamagan, M. Backes, K. Cecire, A. E. Dabrowski, C.
  Darve, J. Ellis, J. A. Gray, E. Kasai, S. Muanza, J. Ndjamba1, A. Philander,
  M. Shahungu, G. Simon, D. Singh, R. Steenkamp, R. Voss, A. Zulu}, {Activity
  Report on the Fifth Biennial African School of Fundamental Physics and
  Applications},
  \url{https://www.africanschoolofphysics.org/wp-content/uploads/2019/08/ASP2018.pdf}
  (2018).

\bibitem{acp2021}
{K\'et\'evi A. Assamagan, Mohamed Chabab, Farida Fassi, Ulrich Goerlach,
  Mohamed Gouighri, et al.}, {The second African Conference on Fundamental and
  Applied Physics}, \url{https://indico.cern.ch/event/1060503/} (2022).

\bibitem{marrakesh}
{Cadi Ayyad University}, \url{https://www.uca.ma/}.

\bibitem{rabat}
{Mohammed V University}, \url{http://www.um5.ac.ma/um5/}.

\bibitem{AfPS}
{The African Physical Society}, \url{https://www.africanphysicalsociety.org/}.

\bibitem{asfap}
{K\'et\'evi A. Assamagan, Simon H. Connell, Farida Fassi, Fairouz Malek,
  Shaaban I. Khalil, et al.}, {The African Strategy for Fundamental and Applied
  Physics}, \url{https://africanphysicsstrategy.org/} (2021).

\bibitem{aas}
{The African Academy of Sciences}, \url{https://www.aasciences.africa/}.

\bibitem{unesco}
{The United Nations Educational, Scientific and Cultural Organization},
  \url{https://en.unesco.org/}.

\bibitem{iaea}
{International Atomic Energy Agency}, \url{https://www.iaea.org/}.

\bibitem{cambridge}
{University of Cambridge}, \url{https://www.cam.ac.uk/}.

\bibitem{psi}
{Paul Scherrer Institute}, \url{https://www.psi.ch/en}.

\bibitem{desy}
{Deutsches Elektronen-Synchrotron}, \url{https://www.desy.de/}.

\bibitem{ucad}
{Universit\'e Cheikh Anta Diop de Dakar}, \url{https://www.ucad.sn/}.

\bibitem{ICTP}
{The International Center for Theoretical Physics}, \url{https://www.ictp.it/}.

\bibitem{IIP}
{Investing In People (IIP) ASBL}, \url{www.semainedelasciencerdc.org}.

\end{thebibliography}

\end{document}